\documentclass[ trackchanges]{aastex7}

\usepackage{hyperref}

\begin{document}

\title{Deep learning-driven atmospheric parameter prediction for hot subdwarf stars with synthetic and observed spectra}

\correspondingauthor{Zhenxin Lei}
\email{leizhenxin@xtu.edu.cn}

\author[0000-0003-2362-6607]{Zhenxin Lei}
\affiliation{Key Laboratory of Stars and Interstellar Medium, 
Xiangtan University, Xiangtan 411105, People's Republic of China}
\email{}
  
\author{Yangyang Dong}
\affiliation{Key Laboratory of Stars and Interstellar Medium, 
Xiangtan University, Xiangtan 411105, People's Republic of China}
\email{}

\author{Bokai Kou}
\affiliation{Key Laboratory of Stars and Interstellar Medium, 
Xiangtan University, Xiangtan 411105, People's Republic of China}
\email{}

\author{Mengqi Feng}
\affiliation{Key Laboratory of Stars and Interstellar Medium, 
Xiangtan University, Xiangtan 411105, People's Republic of China}
\email{}

\author[0000-0002-1599-1519]{Ke Hu}
\affiliation{Key Laboratory of Stars and Interstellar Medium, 
Xiangtan University, Xiangtan 411105, People's Republic of China}
\email{}

\author[0000-0002-9474-4374]{Yude Bu}
\affiliation{School of Mathematics and Statistics, Shandong University, Weihai 264209, Shandong, People's Republic of China}
\email{}

\author[0000-0003-2868-8276]{Jingkun Zhao}
\affiliation{Key Laboratory of Optical Astronomy, National Astronomical Observatories, Chinese Academy of Sciences, Beijing 100012, China}
\email{}

\begin{abstract}
We design a convolutional  neural network (CNN) incorporating channel attention and spatial attention mechanisms to predict  atmospheric parameters of hot subdwarfs. The experimental dataset comprises spectra at nine distinct signal-to-noise ratio (SNR) levels, with each SNR level containing 11 396 synthetic  spectra and 945 observed spectra. The trained deep learning  models achieves mean absolute errors (AME) in predicting hot subdwarf atmospheric parameters of 730 K for effective temperature ($T_{\rm eff}$), 0.09 dex for surface gravity (${\rm log}\ g$), and 0.03 dex for helium abundance ($\log(n{\rm He}/n{\rm H})$), respectively, which ‌reaches the accuracy of traditional spectral fitting methods. Utilizing the trained deep learning models and low-resolution spectra from LAMOST DR12, we confirm 1512 hot subdwarfs from the catalog of hot subdwarf candidates, of which 291 are newly identified. Our results demonstrate that the deep learning model ‌not only‌ achieves accuracy comparable to traditional methods in obtaining hot subdwarf atmospheric parameters,  ‌but also‌ ‌far exceeds‌ them in speed and efficiency, making it ‌particularly suitable for the analysis of large datasets of hot subdwarf spectra.

\end{abstract}


\section{Introduction} \label{sec:intru}
Hot subdwarfs are a highly distinctive class of celestial objects in the Milky Way, spectroscopically classified as type B (sdB) and type O (sdO). These stars occupy late evolutionary stages characterized by ongoing helium (He)-burning in their cores and enveloped by an extremely thin hydrogen shell (e.g., $M_{\rm en}$ $\leq0.01$ $M_\odot$, \citealt{2009ARA&A..47..211H, 2016PASP..128h2001H}). Some of them even evolve off core He-burning stage and toward to white dwarf (WD) cooling curve.  According to the spectral line characteristics, hot subdwarfs can be further subdivided into six subclasses: sdB, sdOB, sdO, He-sdB, He-sdOB, and He-sdO \citep{1990A&AS...86...53M, 2017A&A...600A..50G, 2018ApJ...868...70L}. Furthermore, 
\citet{2013A&A...551A..31D} proposed a new three-dimensional MK-like classification\footnote{Morgan-Keenan (MK) classification system is a two-dimension spectral classification for regular stars based on stellar temperature and luminosity \citep{1943assw.book.....M, 1973ARA&A..11...29M}. While the MK-like classification is a three-dimensional spectral system that extends the two-dimensional MK classification by adding helium abundance as an extra dimension. It was specifically proposed by \citet{2013A&A...551A..31D} for the spectral classification of hot subdwarf stars.}  method for hot subdwarf stars (also see \citealt{2024PASJ...76.1084Z, 2021MNRAS.501..623J}), which gives proxies for effective temperature (spectral class), surface gravity (luminosity class), and surface He abundance (He class) in the classification.
Despite their relatively low masses of approximately 0.5 $M_\odot$ , they exhibit remarkably high surface effective temperatures ($T_{\rm eff}$,  exceeding 20\ 000 K) and high surface gravity (${\rm log}\ g$, exceeding 5.0). Hot subdwarfs in globular clusters occupy the bluest end of the horizontal branch, they are thus also known as extreme horizontal branch (EHB) stars \citep{2009Ap&SS.320..261C}.

The astrophysical significance of hot subdwarfs spans multiple frontier research domains. First, investigating their formation mechanisms advances our understanding of low-mass stellar structure and evolution, and provides critical constraints on binary evolution processes \citep{2012ApJ...746..186C, 2022ApJ...933..137G, 2024ApJ...961..202G, 2025MNRAS.539.3273R}. Second, their surfaces display substantial chemical abundance variations attributed to element diffusion processes, establishing them as ideal laboratories for studying stellar interior physics \citep{2014A&A...565A.100M, 2018MNRAS.475.4728B, 2018ApJ...863...12L}. Third, asteroseismology applications to pulsating subdwarfs enable precise determinations of internal structures and physical parameters to test stellar evolution theories \citep{2010MNRAS.409.1509K, 2018ApJ...853...98Z}. Finally, compact binary systems comprising hot subdwarfs and white dwarfs serve as prominent gravitational wave sources and potential Type Ia supernova progenitors \citep{2018MNRAS.480..302K, 2024ApJ...963..100K, 2009MNRAS.395..847W}. 

The formation mechanism of hot subdwarfs remains unclear. Since over half of the observed hot subdwarfs reside in close binary systems \citep{2001MNRAS.326.1391M,2004Ap&SS.291..321N,2011MNRAS.415.1381C, 2025A&A...693A.121H}, binary evolution is usually considered as the dominant channel for their formation. \citet{2002MNRAS.336..449H, 2003MNRAS.341..669H} used binary population synthesis to demonstrate that sdB stars can be formed through three pathways: stable Roche-lobe overflow (RLOF), common envelope (CE) ejection, and the merger of two He-WDs. \citet{2017ApJ...835..242Z} investigated the evolution after the merger of a He WD with a low-mass main-sequence star, revealing that this process could produce intermediate He-rich and single hot subdwarfs (also see \citealt{2012MNRAS.419..452Z}). Additionally, \citet{2021MNRAS.507.4603M} studied the stripping of the companion envelope during Type Ia supernova explosions from massive carbon-oxygen WDs accreting matter from main-sequence stars, which could also generate intermediate He-rich hot subdwarfs (also see \citealt{2024RAA....24e5003J, 2020ApJ...903..100M}). \citet{2024ApJ...964...22L} found that He-rich massive hot subdwarfs may originate from CE  ejection during the asymptotic giant branch (AGB) phase. This model could explain the discovery by \citet{2022MNRAS.515.3370L} of a short-period, massive sdO/B binary system. The strong Ca H\&K lines with large blue shift ($\sim$ 200 m/s) observed in this system suggest it has just undergone the CE  ejection process.

Observationally, significant progress has been made in hot subdwarf studies. \citet{2011A&A...530A..28G} identified over 1100 hot subdwarfs in SDSS survey data, while \citet{2019MNRAS.486.2169K} confirmed 415 hot subdwarfs in SDSS Data Release 14 (DR14). Some researchers utilized the  Large Sky Area Multi-Object Fiber Spectroscopic Telescope (LAMOST, \citealt{2012RAA....12.1197C,2012RAA....12..723Z,2006ChJAA...6..265Z}) to identify nearly 2000 hot subdwarfs, obtaining precise atmospheric parameters \citep{2018ApJ...868...70L, 2019ApJ...881..135L,2020ApJ...889..117L, 2023ApJ...942..109L, 2019ApJ...881....7L,2021ApJS..256...28L}. They also derived spatial positions, velocity vectors, and orbital parameters for these stars using parallax and proper motion data from Gaia Early Data Release 3 \citep{2021A&A...649A...1G}. Based on these observationally confirmed hot subdwarfs, \citet{2023ApJ...942..109L} determined the masses, radii, and luminosities of 664 such stars by comparing observed fluxes with synthetic  spectral energy distributions (SEDs). Meanwhile, \citet{2024ApJS..271...21L} obtained carbon and nitrogen abundances for 210 He-rich hot subdwarfs observed by LAMOST. These observational results provide essential parameters for subsequent studies of hot subdwarfs. 

Additionally, artificial intelligence methods have also achieved significant progress in searching for hot subdwarfs. \citet{2017ApJS..233....2B} firstly used the hierarchical extreme learning machine (HELM) algorithm to identify over 7000 hot subdwarf candidates in LAMOST Data Release 1 (DR1). \citet{2019PASJ...71...41L} subsequently confirmed 56 hot subdwarfs from these candidates. After that, \citet{2019ApJ...886..128B} employed a convolutional  neural network (CNN) combined with support vector machines (SVM) to discover 784 hot subdwarf candidates in LAMOST DR4, of which 207 were confirmed as hot subdwarfs. Subsequently, \cite{2022ApJS..259....5T} utilized a two-step classification framework integrating binary and eight-class CNNs, identifying 2393 hot subdwarf candidates in LAMOST DR7. Among these candidates, 2067 were confirmed as hot subdwarfs, including 25 newly identified ones. \citet{2024ApJS..274....2C} applied a Squeeze-and-Excitation Residual Network (Se-ResNet) and SVM model to find 3086 hot subdwarf candidates in LAMOST DR8, confirming 168 new hot subdwarfs. \citet{2025A&A...693A.245W} used deep learning models to identify 3413 high-confidence hot subdwarf candidates within the SDSS image dataset and confirmed 331 new hot subdwarfs. 

With the sequential release of observational data from  new generation of large-scale sky survey telescopes, such as: Gaia DR3 \citep{2023A&A...674A...1G}, SDSS\citep{2022ApJS..259...35A}, LAMOST \citep{2012RAA....12.1197C}, TESS \citep{2014SPIE.9143E..20R}, DESI\citep{2016arXiv161100036D}, etc, newly discovered hot subdwarfs will experience a dramatic increase. However, measurement of their atmospheric parameters still relies on traditional methods. The conventional approach for determining atmospheric parameters of hot subdwarfs involves spectral fitting, where observed spectra are matched against synthetic spectra models. The parameters corresponding to the best-fitting synthetic spectrum are then assigned to the observed spectrum. While this method delivers reliable results, it is computationally intensive and time-consuming. The process requires both the generation of synthetic spectra with non-local thermodynamic equilibrium (N-LTE) atmospheres and the iterative search for optimal model fits. For instance, fitting a single LAMOST low-resolution spectrum using the   $\chi^{2}$ minimizing fitting program {\sc XTgrid} typically requires approximately 5 to 30 minutes of computation time (\citealt{2012MNRAS.427.2180N, 2018ApJ...868...70L}, and references therein). Therefore, this traditional method suffers from limitations such as low efficiency, slow speed, and high dependence on spectrum  quality, making it impractical for large-scale spectral data analysis. 

In this paper, we employ a CNN model incorporating channel attention and spatial attention mechanisms to predict the atmospheric parameters of hot subdwarfs. This deep learning model utilizes both synthetic spectra augmented with noise and observed spectra as training samples, mitigating the limitations and imbalances associated with using only observed spectra. We also trained nine distinct deep learning models tailored to observed spectra with varying signal-to-noise ratios (SNRs) for atmospheric parameter prediction. The structure of the paper is as follows: Section 2 describes the architecture of the CNN,  the hyperparameter used during training, and the construction of the training dataset. Section 3 presents the results of the deep learning models, compares the predicted parameters with those obtained from spectral fitting method, and utilizes the trained models to identify 291 new hot subdwarfs. Finally, Section 4 concludes with a brief discussion of the model and a summary of the work. 

\section{convolutional  neural network and  training spectra } \label{sec:spectra}
\subsection{Deep learning model architecture}

\begin{deluxetable}{c|c|c|c|c|c|c}
\tablenum{1}
\tablecaption{The structure of convolutional  neutral network adopted in this study. Training data flow in CNN architecture from input layer (top) to final dense layer ( bottom ). The total trainable and non-trainable parameters are 389 600 and 0, respectively.}
\tabletypesize{\normalem}
\tablehead{ \colhead{Layer name} \vline &  \colhead{Filters} \vline &  \colhead{Kernel/Pool  size} \vline & 
\colhead{Activation}  \vline &\colhead{Output shape} \vline & \colhead{Trainable parameters} \vline }
\startdata 
Input Layer & &&&(None, 2200,1) & 0\\
\hline
Convolutional layer & 32& 8& ReLU&  (None, 2200, 32)& 288\\  
\hline
Max pooling layer& & 4& & (None, 550, 32) & 0     \\ 
\hline
Convolutional layer& 64 & 6 & ReLU &(None, 550, 64) & 12 353 \\
\hline
Channel attention layer&&&&(None, 550, 64)& 1354\\
\hline
Spatial attention layer&&&&(None, 550, 64)& 15\\
\hline
Max pooling layer& & 4& & (None, 137,64) & 0\\
\hline
Convolutional layer& 128 & 5 & ReLU &(None, 137, 128) & 41 088\\
\hline
Channel attention layer &&&&(None, 137, 128)& 5525\\
\hline
Spatial attention layer&&&&(None, 137, 128)& 15\\
\hline
Max pooling layer& & 4& & (None, 34,128) & 0\\
\hline
Convolutional layer & 256 & 4 & ReLU &(None, 34, 256) & 131 328\\
\hline
Global average pooling layer &&&&(None, 256)&0 \\
\hline
Dense Layer &&&ReLU&(None, 512)&131 584\\
\hline
Dropout layer &&&&(None, 512)&0\\
\hline
Dense Layer &&&ReLU&(None, 128)&65 664\\
\hline
Dense Layer &&&&(None, 3)&387\\
\enddata
\end{deluxetable}

 Using the TensorFlow deep learning framework \citep{abadi2016tensorflowlargescalemachinelearning}, a convolutional neural network (CNN) was constructed. This model achieves high-precision prediction of hot subdwarf atmospheric parameters through training both on synthetic and observed spectra of hot subdwarfs. Table 1 shows the structure of the network. It consists of four convolutional  layers, three max pooling layers, one global average pooling layer and three fully connected (dense)  layers.   Convolutional  layer conduct the convolution  operation between input data and corresponding weights \citep{1998IEEEP..86.2278L, 2015Natur.521..436L, 2018PhLB..778...64G}. In this layer, each neuron in output layer is connected to local regions of the input layer, and connections between input and output have different weights. Typically, a convolutional layer is often followed by a pooling layer. Pooling layer performs ‌downsampling‌ on the output of the convolutional layer and preserves the useful information while reducing the size of the data. This layer could reduce the height and width of the input, and helps to reduce computation time. 

An channel attention layer \citep{hu2019squeezeandexcitationnetworks} and a spatial attention layer \citep{woo2018cbamconvolutionalblockattention} follow closely behind the second and third convolutional  layers (see Table 1), respectively. The channel attention layer could assess the contribution level of different feature channels (such as multi-level features extracted by convolutional  kernels). For example, certain channels may encode temperature-sensitive features, while others capture surface gravity information. On the other hand, spatial attention would focus on the importance variation across different wavelength positions in spectral data. For instance, it boosts the weights of key regions containing specific spectral lines (such as hydrogen or helium lines) that are critical for parameter prediction. Three fully connected layers are behind the global average pooling layer. The first 512-unit fully connected layer will consolidate locally extracted spectral features from convolutional  layers into global representations, and learn higher-order nonlinear relationships through activation functions. The second 128-unit fully connected layer will further reduce feature dimensions to enhance model generalization ability, while the last 3-unit layer will directly corresponds to prediction of the three atmospheric parameters (e.g., $T_\mathrm{eff}$, $\mathrm{log}\ g$ and $\log(n{\rm He}/n{\rm H})$). 

For each convolutional  and fully connected layer, we applied the Rectified Linear Unit (ReLU) activation function and used the same padding method. This enables the model to fit complex nonlinear relationships within the spectral data and preserves the structural integrity of the spectra.

During model training, we employed the Adam optimizer with its adaptive learning rate. Adam dynamically adjusts the learning rate size based on gradient changes during training, providing an automatic and efficient optimization path that significantly enhances model performance and training efficiency. The total number of training epochs is set to 300. We utilized a callback function to monitor the loss function on the validation data. During training, automatic early stopping will trigger if the validation loss exhibits minimal improvement over consecutive epochs. The model with the smallest loss function observed during training was automatically saved. This approach prevents ineffective iterations and mitigates overfitting.

\subsection{ Training sample: synthetic spectra}

According to the recent catalog of \citet{2022A&A...662A..40C}, there are more than 6000 hot subdwarf stars having been identified, but only more than 1000 stars having LAMOST spectra with good quality (e.g., SNR $\geq$ 10). Therefore, it is insufficient quantity if only using observed spectra as training sample in the deep learning model. Furthermore, the numbers of known hot subdwarf stars with variety of spectral types are very different from each other, e.g., sdB/sdOB stars accounts for  majority of the stellar population, while other stars (e.g., He-sdB, He-sdO, He-sdOB) are much less. The lack of total observed spectra and obvious imbalance in the number counts of training sample with various parameters would influence the final prediction results significantly.  

\begin{figure}
    \centering    
    \includegraphics[width=170mm]{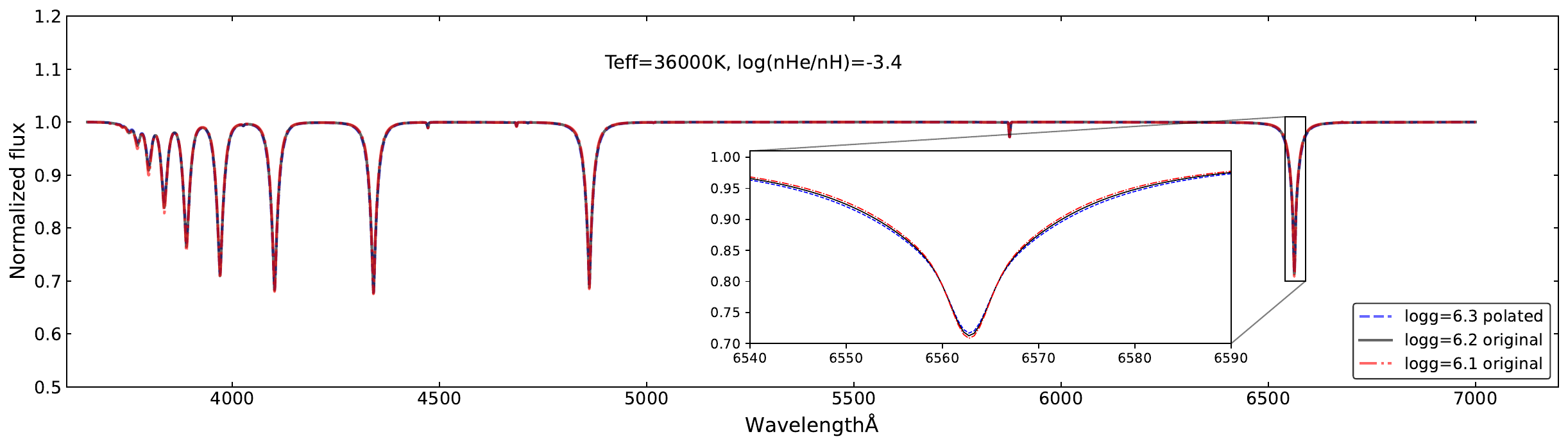}
    \caption{A synthetic spectrum of ${\rm log}\ g$ = 6.3 (blue-dashed curve) obtained by extrapolation from the synthetic spectrum with ${\rm log}\ g$ = 6.2 (gray-solid curve) and 6.1 (red dot-dashed curve). The three synthetic spectra shown in the figure have same surface temperature ($T_{\rm eff}$ = 36\ 000 K) and He abundance ($\log(n{\rm He}/n{\rm H})$ = -3.40).  The inset plot magnifies the region surrounding the $H_\alpha$ line to demonstrate the subtle variations among the three spectra with little   different ${\rm log}\ g$ values. See the context for details.}   
    \label{fig syn_spec_polate}
\end{figure}

In order to solve the problems mentioned above, we use  synthetic spectra of hot subdwarf stars calculated in \citet{2014ASPC..481...95N} as the main training sample. The synthetic spectra in \citet{2014ASPC..481...95N} were calculated with Non-local thermodynamic equilibrium (N-LTE) model atmosphere code {\sc Tlusty} 204 and {\sc Synspec} 49 \citep{1995ApJ...439..875H, 1995ApJ...439..905L}. The grid of the spectra basically covers the observed parameter space of hot subdwarf stars, e.g., 20 000K $\leq$ $T_\mathrm{eff}$  $\leq$ 56 000K with step size of 1000K, 5.0 $\leq$ $\mathrm{log}\ g$  $\leq$ 6.3 with step size of 0.1 dex and -4.3 $\leq$  $\log(n{\rm He}/n{\rm H})$ $\leq$ 2.0 with step size of about 0.03 dex (see Table 1 in their study). It consists of 11 396 grid points to which 10 887 spectra were converged in the calculation. These spectra are in the wavelength range of 3200 - 7200\AA \ with 0.1\AA \ step size. 

We complemented the  509 missing spectra through linear extrapolation method using the already existed spectra. Fig \ref{fig syn_spec_polate} displays an example of  a synthetic spectrum for ${\rm log}\ g$ = 6.3 (blue-dashed curve) ‌produced by extrapolating‌ the spectra of ${\rm log}\ g$ = 6.1 and ${\rm log}\ g$ = 6.2, all with identical effective temperature and helium abundance. The ${\rm log}\ g$ values of the three spectra are relatively close, resulting in minor differences between the corresponding spectra. Nevertheless, we can observe distinctions among the three spectra in $H_\alpha$ line region within the inset plot. The spectrum with the highest gravitational acceleration (e.g. ${\rm log}\ g$ = 6.3, blue-dashed curve) exhibits the broadest line wings and shallowest line depth. In contrast, the spectrum with the lowest gravitational acceleration (e.g. ${\rm log}\ g$ = 6.1, red dot-dashed curve) shows the narrowest line wings and deepest line depth. The spectrum with ${\rm log}\ g$ = 6.2 (gray-solid curve) demonstrates line widths and depths between these two extremes. These features could be captured by the CNN models.

It should be emphasized that although spectra supplemented by linear extrapolation may lack precision, they have minimal impact on model training. Among the 509 missing synthetic spectra in \citet{2014ASPC..481...95N}, most correspond to high surface gravities (e.g., ${\rm log}\ g$ = 6.2 – 6.3). In observed low-temperature hot subdwarfs (e.g., sdB/sdOB, $T_\mathrm{eff} \le$ 35\ 000 K), surface gravity typically remains below 6.0. Thus, spectra exhibiting both low temperature and high gravity are observationally absent. Conversely, high-temperature subdwarfs (e.g., sdO, $T_\mathrm{eff} \geq $ 40\ 000 K) do show cases with ${\rm log}\ g \geq$  6.0. Crucially, our training sample also incorporates observed spectra (see Sect 2.3), which include many high-temperature and high-${\rm log }\ g$ samples. It  significantly enhances model accuracy for such subtypes. Therefore, the 509 missing spectra are non-critical for model performance. Nevertheless, to ensure sample balance, we included them in the training set.‌ Thus, 11 396 synthetic spectra have been generated as the training sample, and it is balanced for spectra with different parameters. 

Since our goal is to derive the atmospheric parameters of hot subdwarfs with LAMOST spectra, the resolution of these synthetic spectra is degraded to match that of the LAMOST spectra (e.g., $\frac{\lambda}{\Delta \lambda}$ = 1800) before training. 
The key differences between observed and synthetic spectra are the presence of noise that observed spectra contain noise while synthetic spectra do not. Therefore, if we use synthetic spectra as training sample for deep learning models, artificial noise needs to be added to them. Given the inherent SNR variations across observed spectra, we must generate synthetic spectra spanning multiple SNR levels. This enables training SNR-specialized deep learning models capable of robust atmospheric parameter prediction from observed spectra at any noise level. 

 LAMOST spectra exhibit different SNRs across the u, g, and r bands (typically lowest in u band, highest in g band, and intermediate in r band), thus we set SNRg = 1.9 $\times$ SNRu and SNRr = 1.5 $\times$ SNRu for each corresponding synthetic spectrum. Considering that each observed spectrum has a different SNRu, our synthetic spectra need to cover the entire SNRu range of the observed spectra. However, since there are 11 396 synthetic spectra corresponding to each SNRu value, generating spectra for continuous SNRu values would result in an excessively large training set. Therefore, we selected synthetic spectra at ‌non-continuous‌ SNRu values for training. Considering that the SNRu of spectra selected from the LAMOST database ranges from 8.0 to several hundred (see Section 3.1) and peaking around 15.0, we generated synthetic spectra with nine SNRs in the u band, specifically SNRu = 15, 20, 25, 35, 45, 60, 80, 100 and 200, respectively. Considering that spectral quality improves with higher SNRu, the interval gradually increases. This approach not only effectively reduces the total number of synthetic spectra, enabling efficient training process, but also ensures no significant impact on the prediction of atmospheric parameters (see the discussions in Sect 4 for model dependence on SNRu). 

\begin{figure}
    \centering    
    \includegraphics[width=160mm]{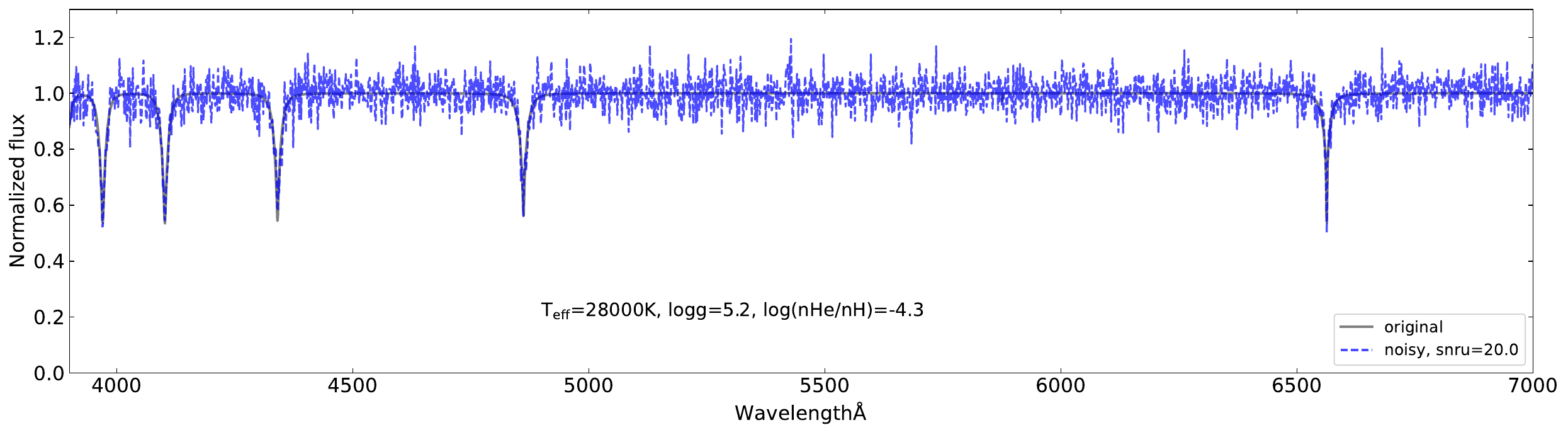}\\
    \includegraphics[width=160mm]
    {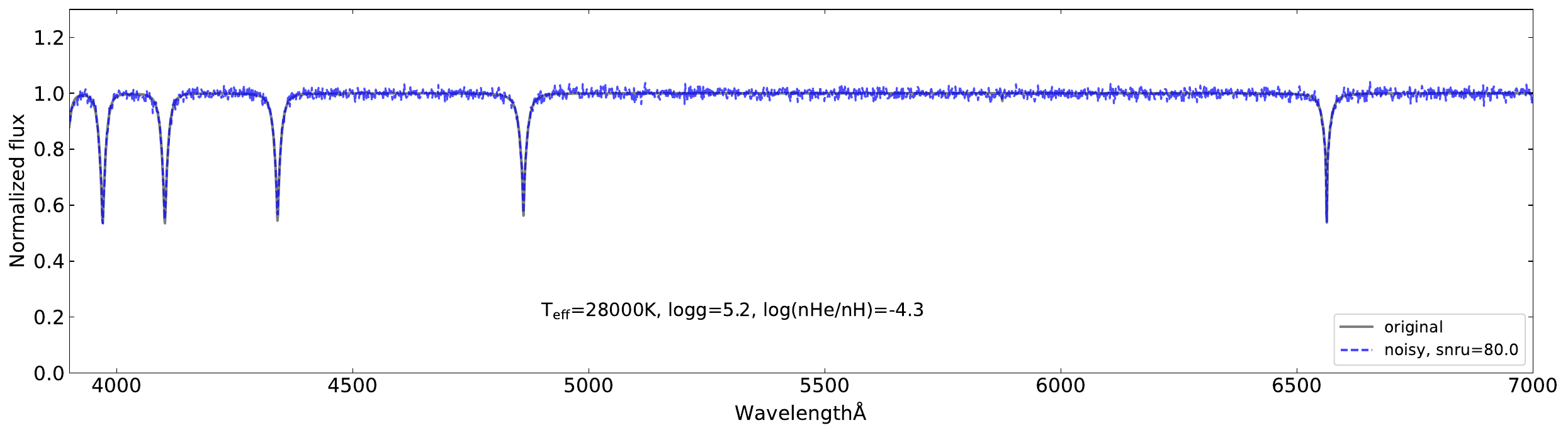}

    \caption{Top panel: a synthetic spectrum without noise (black-solid curve) and with noise (blue-dashed curve) of SNRu=20.0. Bottom panel: the same synthetic spectrum without noise (black-solid curve) and with noise (blue-dashed curve) of SNRu=80.0.} 
    \label{fig spectrum_noisy}
\end{figure}

For a given pass-band, such as the u band, we used the designated input average SNRu (e.g., SNRu = 20) as the mean value, and 0.1 $\times$ SNRu  as the standard deviation. ‌We assigned the SNRu at each individual sampling point of spectrum by drawing from this Gaussian distribution.‌ Subsequently, the flux error (i.e., $\sigma_{\rm flux}$) per sampling point was calculated as $\sigma_{\rm flux}$ = $\frac{\rm flux}{\rm SNRu}$. These operations were also repeated for sampling points in g and r band for the same spectrum. Finally, we generated the synthetic noisy flux value at each point by sampling from a Gaussian distribution where the synthetic flux at that point was the mean value and the derived $\sigma_{\rm flux}$ was the standard deviation. 

‌Using this methodology, we produced a complete set of noise-added synthetic spectra for all nine SNRu levels.‌ Each SNRu level comprises 11 396 individual synthetic spectra.  Figure \ref{fig spectrum_noisy} illustrates a comparison of an sdB-type hot subdwarf spectrum before and after adding noise corresponding to an average SNRu of 20.0 (top panel) and 80.0 (bottom panel), respectively. 

\subsection{ Training sample: observed spectra}
Although we incorporated noise into the synthetic spectra, they still differ to some extent from the actual observed spectra of hot subdwarfs.‌ Additionally, a small number of confirmed hot subdwarfs possess effective temperatures and surface gravities that exceed the parameter coverage of our synthetic spectral grid (e.g., $T_{\rm eff} > 56\ 000$K, ${\rm log}\ g> 6.3$, \citealt{2022A&A...662A..40C}).

‌To improve the accuracy of model predictions, we augmented the training sample by including observed spectra of confirmed hot subdwarfs.‌ These 945 confirmed sources were selected from \citet{2018ApJ...868...70L, 2019ApJ...881..135L, 2020ApJ...889..117L, 2023ApJ...942..109L}, and their corresponding LAMOST spectra are added to the training sample. ‌This set of 945 observed spectra encompasses all six spectral types of hot subdwarfs.‌ For each of the 11 396 synthetic spectra generated at a specific SNRu level, we added these 945 observed spectra (also see the discussion in Section 4).‌ Consequently, ‌each complete training sample subset comprises 12 341 (11 396 synthetic + 945 observed) spectra. Both synthetic and observed spectra are normalized by using the normalization module integrated in the LASPEC package \citep{2020ApJS..246....9Z,  2021ApJS..256...14Z} before training,  and they were interpolated into the wavelength range of 3900 - 7000 \AA \ in which 2200 sampling points were included (see Fig \ref{fig spectrum_noisy}).    

For each set of spectra, it needs to be split into training set, test set and validation set. Training set directly participates in the training process to optimize model parameters, fitting complex mappings between spectral features and atmospheric parameters. Validation set is used for hyperparameter tuning and early stopping to prevent overfitting and enhance model stability. Test set remains entirely independent of training. Upon completion of model training, it evaluates the model's generalization performance.

For the 12 341 training spectra, we partitioned them into training and test sets using an 8:2 ratio. Within the training set, we further divided the training data and validation data using an 8:2 ratio. It  means that for spectra at each SNR level, we had approximately 7900 spectra in the training set, 2470 spectra in the test set, and 1970 spectra in the validation set. All spectra were randomly shuffled before training. 

\section{results} 

\begin{figure}
    \centering    
    
    \includegraphics[width=180mm]{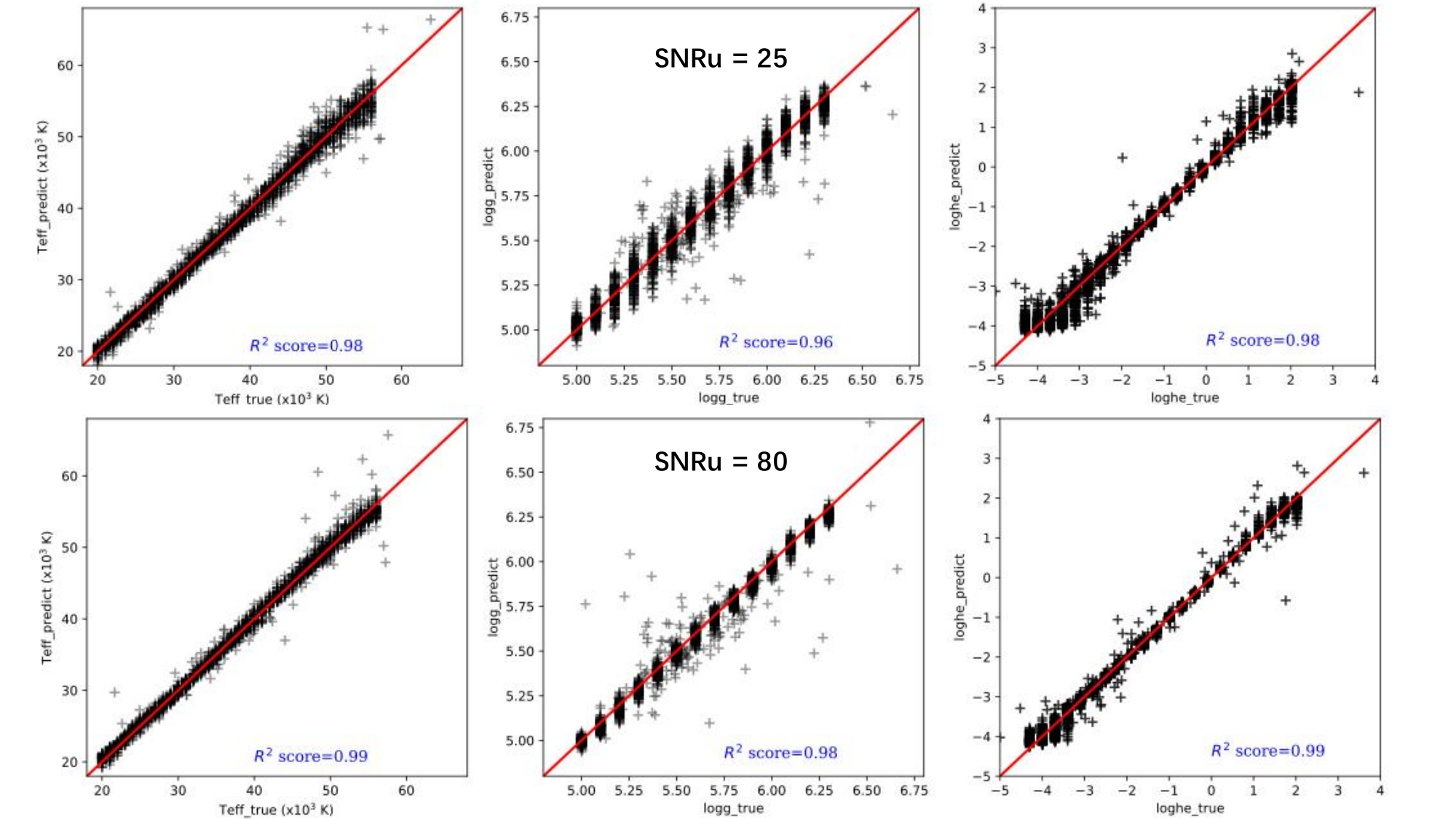}\\
    \caption{Comparisons between CNN model predictions and the true parameter values for the test set. From left to right, it gives the comparison of $T_{\rm eff}$, ${\rm log}\ g$ and 
    $\log(n{\rm He}/n{\rm H})$, respectively. Top panels show the results of SNRu = 25 model, while bottom panels show the results of SNRu = 80 model. Horizontal axis presents parameters from spectral fitting method (training labels), while vertical axis presents CNN model predictions. The coefficient of determination $R^{2}$ is shown in each panel (see text for details). }
    \label{fig comparsion for test data}
\end{figure}

Using the the network designed in Section 2.1, we obtained nine distinct models corresponding to spectra with nine different SNRs for predicting the atmospheric parameters of hot subdwarfs. Fig \ref{fig comparsion for test data} presents the prediction results of the trained models for SNRu=25 (top panels) and SNRu=80 (bottom panels) on the test set data. 
The horizontal axis in the panel represents the true atmospheric parameters (i.e., the training labels) of the test set spectra. The vertical axis represents the atmospheric parameters predicted by the trained models. Each panel displays the coefficient of determination $R^{2}$, which quantifies the goodness-of-fit between the predicted and true values. $R^{2}$ is defined as:   
\begin{equation}
    R^{2}=1-\frac{\sum(y_{i}-y_{pi})^{2}}{\sum(y_{i}-\bar{y})^{2}}, 
\end{equation}
where $y_{i}$ is the label (true value) of atmospheric parameter, $y_{pi}$ is the prediction value, while $\bar{y}$ is average value of the labels. An $R^{2}$ value closer to 1 indicates higher prediction accuracy.

Overall, the $R^{2}$ values for all atmospheric parameters exceed 0.96. Predictions for effective temperature ($T_{\rm eff}$, left panels) and helium abundance ($\log(n{\rm He}/n{\rm H})$, right panels) are particularly accurate, with $R^{2}$ values surpassing 0.98. Even for surface gravity (${\rm log}\ g$, middle panels), the SNRu=25 model achieves $R^{2}$=0.96, while the SNRu=80 model reaches $R^{2}$=0.98. These results demonstrate that models trained at different SNRs successfully capture the primary spectral features of hot subdwarfs and deliver reasonably accurate predictions when applied to test data.

It is the fact that ${\rm log}\ g$  primarily governs the broadening of spectral line profiles. Higher surface gravity induces greater broadening in line profiles such as the Balmer series. However, other factors including stellar rotation, turbulence, and instrumental effects also could contribute to line broadening. Consequently, among the three fundamental atmospheric parameters of hot subdwarfs, ‌${\rm log}\ g$ exhibits the lowest sensitivity to spectral line morphology‌. This inherent limitation leads to its ‌largest measurement uncertainty‌ when determined via synthetic  spectral fitting. In this study, since the training labels for ${\rm log}\ g$  are themselves derived from spectroscopically fitted values, the predicted ${\rm log}\ g$  from our CNN model consequently manifests the highest errors compared to $T_{\rm eff}$ and helium abundance.   

The surface helium abundance of hot subdwarfs varies greatly, with the atomic number ratio $n{\rm He}/n{\rm H}$ ranging from $10^{-5}$ to over 200. This data distribution is extremely asymmetric and imbalanced. Using this data as the training label directly would affect the accuracy of the final results. Therefore, we took the logarithm of the data during training, using $\log(n{\rm He}/n{\rm H})$ as the training label. This made the range of helium abundance values more balanced, from -5 to 2.0, thus improving the model prediction accuracy. However, when dealing with extremely He-poor hot subdwarfs (such as $\log(n{\rm He}/n{\rm H})$ = -4.2 and $\log(n{\rm He}/n{\rm H})$ = -4.5), their spectra either lack helium lines or have very weak helium lines. Combined with the interference of noise, CNNs find it difficult to distinguish the subtle differences in their spectral lines. As a result, the prediction accuracy of helium abundance at the boundaries is not as good as that of intermediate values. 

\subsection{Comparison with the results from spectral fitting}
To further validate the model's generalization capability, we  utilized our trained model to predict the atmospheric parameters of hot subdwarfs with LAMOST spectra from the catalog of \citet{2022A&A...662A..40C} and compared them with existing reference measurements. Initially, we cross-matched the spectroscopically confirmed hot subdwarfs in the catalog of \citet{2022A&A...662A..40C} with the LAMOST DR12 spectral dataset and yielded 2599 spectra. To ensure spectra quality for subsequent analysis, we select LAMOST spectra with SNRu higher than 8.0. After removing duplicated resources, 1534 spectra exhibit a SNR greater than 8.0 in the u band, making them suitable for predicting atmospheric parameters using our pre-trained deep learning models.
For each selected observed spectrum, we chose the corresponding deep learning model based on its SNRu. For example, if SNRu $\le$ 15.0, the model trained for SNRu = 15.0 is used to predict the atmospheric parameters of the observed spectra. 
If 15.0 $<$ SNRu $\le$  20.0, then the model trained for SNRu = 20 is used to predict the atmospheric parameters. Similarly, if 20.0 $<$ SNRu $\le$ 25.0, the model trained for SNRu = 25 is used, and so forth for other ranges (See discussions in Sect 4 for the model dependence on SNRu ). Using this method, we obtained the atmospheric parameters for all 1534 selected spectra. 

\begin{figure}
    \centering    
    
    \includegraphics[width=85mm]{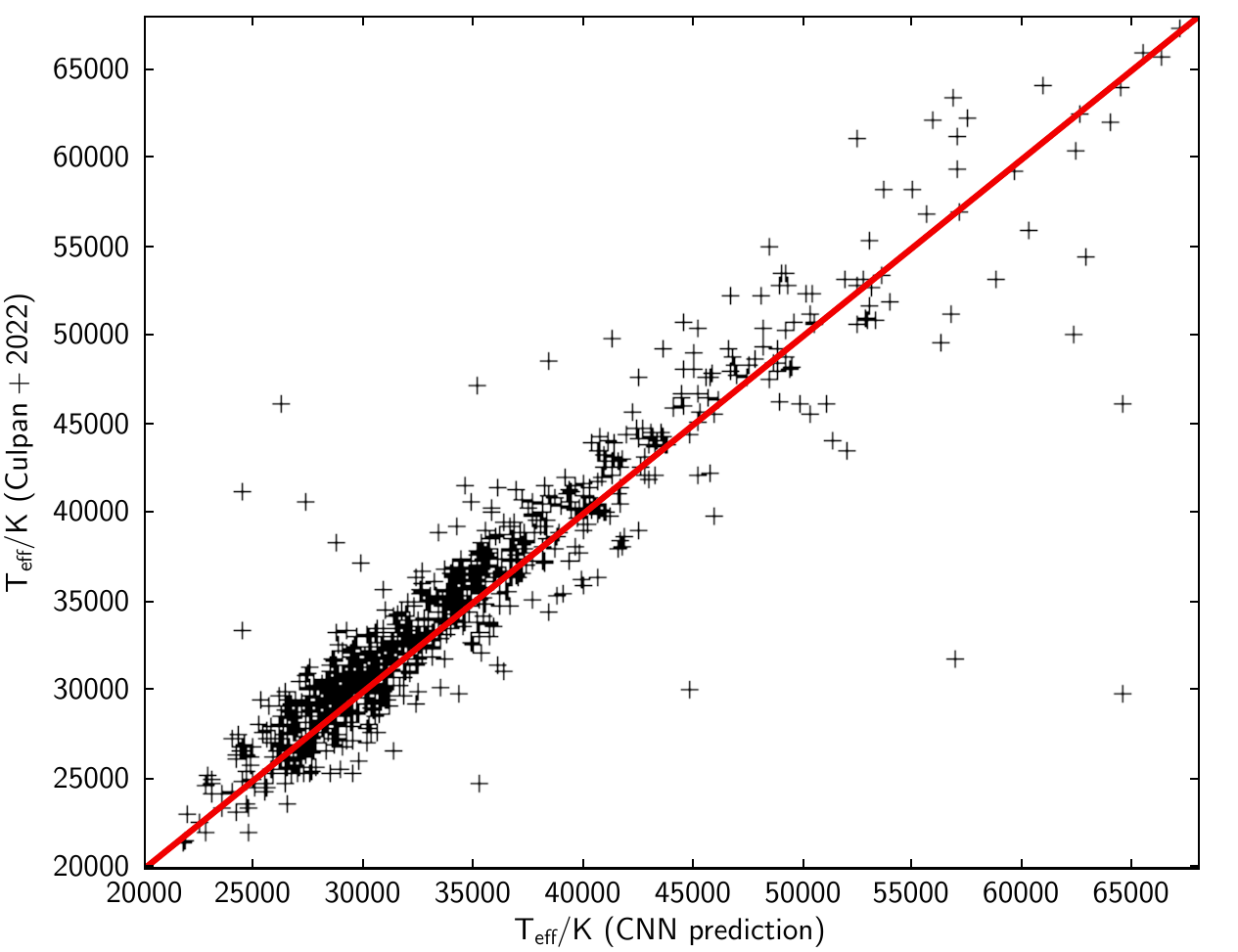}
    \includegraphics[width=85mm]{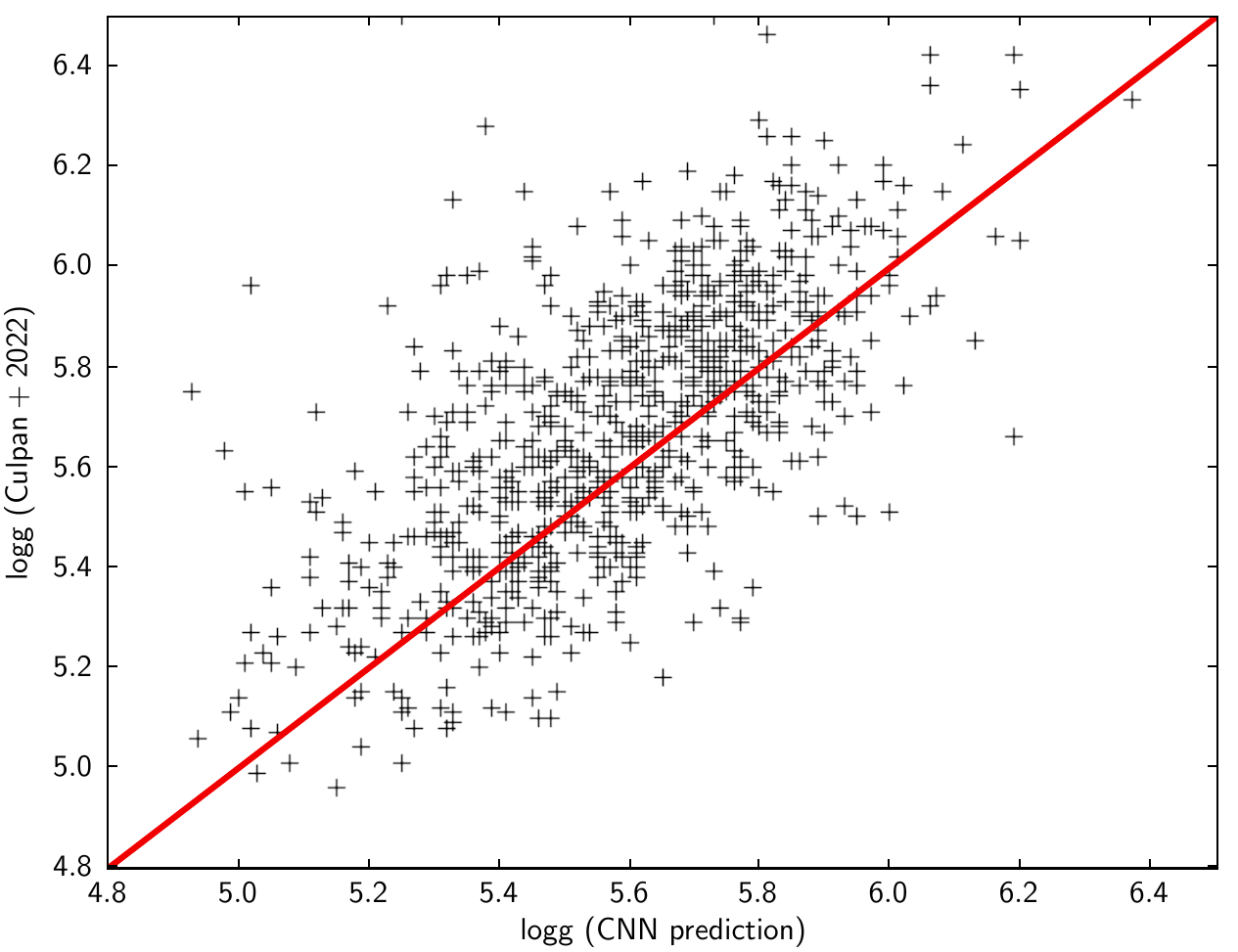}
    \includegraphics[width=85mm]{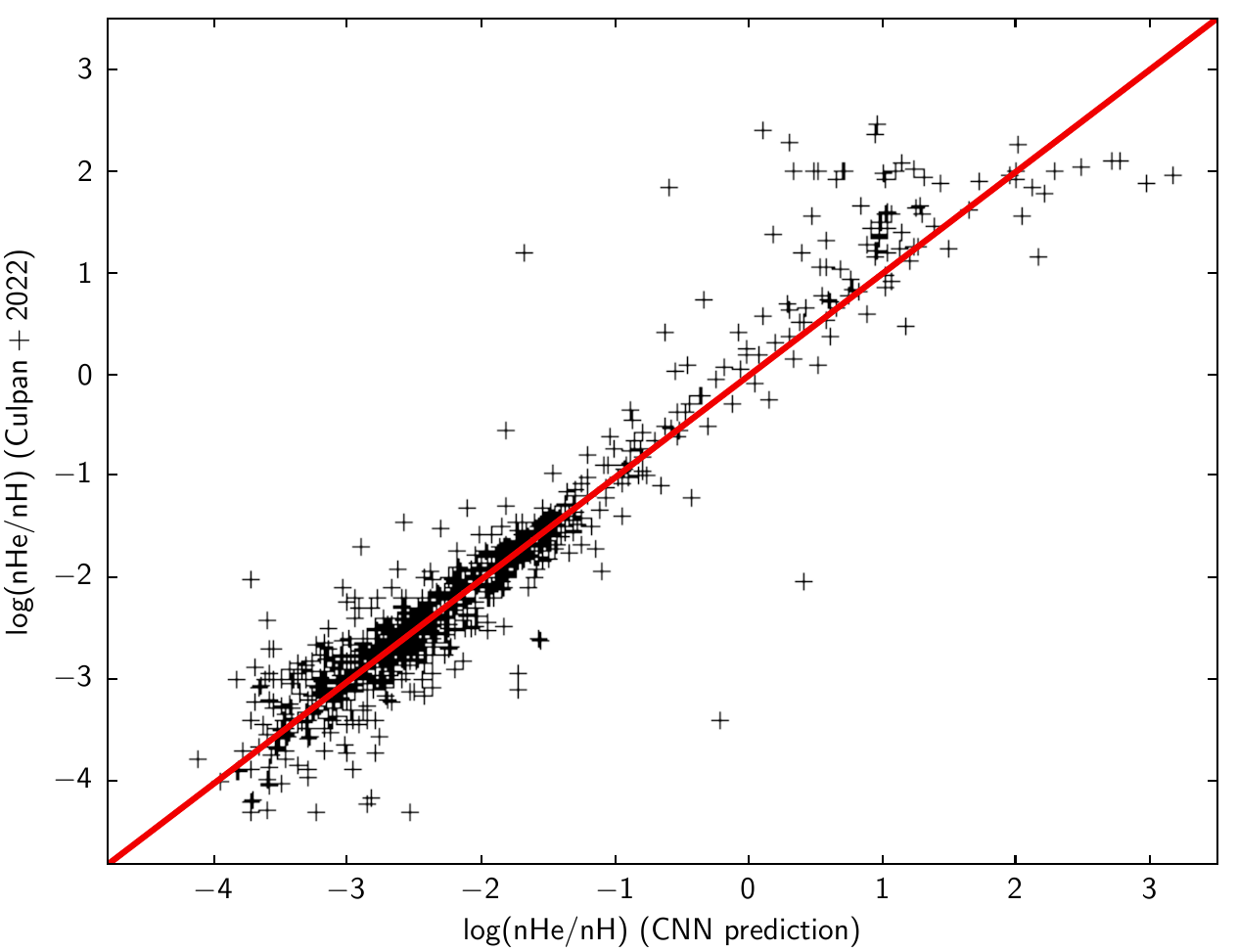}
    \caption{Comparison between CNN model predictions and the parameter values for the known hot subdwarfs in \citet{2022A&A...662A..40C}. The mean absolute errors (MAE) are 730 K for $T_{\rm eff}$ (top left panel), 0.09 dex for ${\rm log}\ g $ (top right panel) and 0.03 dex for $\log(n{\rm He}/n{\rm H})$ (bottom middle panel).  } 
    \label{fig comparison with culpan}
\end{figure}

By cross-matching the selected 1534 stars with the hot subdwarf stars recorded in \citet{2022A&A...662A..40C}, which have detailed atmospheric parameters (i.e, $T_{\rm eff}$, ${\rm log}\ g$, $\log(n{\rm He}/n{\rm H})$), we obtained over 900 common sources\footnote{Note that, not all of the hot subdwarf stars collected in \citet{2022A&A...662A..40C} have measured atmospheric parameters. Specifically, among the 6616 hot subdwarf stars, 2954 have three complete atmospheric parameters, while 3662 are lacking either complete or partial atmospheric parameters.‌ In the comparison of Fig \ref{fig comparison with culpan}, we also removed the stars with $T_{\rm eff} < 20\ 000$K and stars with  ${\rm log}\ g < 5.0$.}. Figure \ref{fig comparison with culpan} shows a comparison between the atmospheric parameters predicted by our deep learning model for these stars and the previously measured values reported in \citet{2022A&A...662A..40C}. Here, the x-axis represents the atmospheric parameters predicted by the deep learning model, while the y-axis represents the parameters previously obtained through methods like spectral fitting. The figure demonstrates excellent consistency between the model-predicted atmospheric parameters and those derived from spectral fitting, with very small mean absolute errors (MAE), i.e., $\Delta$$T_{\rm eff}$ = 730 K, $\Delta$${\rm log}\ g$ = 0.09 dex, and $\Delta$$\log(n{\rm He}/n{\rm H})$ = 0.03 dex. Among the three parameters, although the prediction of ${\rm log}\ g$ exhibits greater uncertainty than the other two ($T_{\rm eff}$ and $\log(n{\rm He}/n{\rm H})$), showing a larger scatter between predicted and measured values, the accuracy achieved by the deep learning model fully matches that attained by spectral fitting. These results confirm that the atmospheric parameters predicted by our deep learning models are highly accurate and reliable.

\subsection{New hot subdwarf stars identified with CNN models.}
In the work of \citet{2022A&A...662A..40C}, besides the catalog of 6616 confirmed hot subdwarfs, they also selected 61\ 585 high-confidence hot subdwarf candidates by combining Gaia photometry and proper motions. We cross-matched this candidate catalog with the LAMOST DR12 dataset, selecting sources with SNRu larger than 8.0, which yielded spectra for 3277 candidate objects. After removing duplicates, spectra from 2013 unique candidates were input into our deep learning models for atmospheric parameter prediction. 

Prior to inputting the spectra into the trained deep learning models, we required measuring the radial velocity (RV) of each spectrum and applying an RV correction to its wavelength scale. In this work, we utilized the {\sc Laspec} software package \citep{2021ApJS..256...14Z,2020ApJS..246....9Z} to measure the radial velocities for the LAMOST spectra. {\sc Laspec} employs the cross-correlation function (CCF) method for RV determination \citep{1979AJ.....84.1511T, 2015AJ....150..173N, 2020AJ....160...82S}. When computing radial velocities, we employed a radial velocity grid ranging from -1000 to 1000 km/s with a step size of 10 km/s. The associated uncertainties were determined through a Monte Carlo method, which provided radial velocity values corresponding to the 16th, 50th, and 84th percentiles of the confidence interval. From these results, we derived the corresponding asymmetric error bars. Detailed methodology can be found in Section 3.3 of \citet{2021ApJS..256...14Z}. As our targets are hot subdwarfs, we selected 700 synthetic hot subdwarf spectra from \citet{2014ASPC..481...95N} to serve as templates for the RV measurements. These template spectra comprehensively cover the atmospheric parameter space of hot subdwarfs and include all known spectral subtypes.‌

\begin{deluxetable}{lcllllll}
\tablenum{2}
\tablecaption{The main parameters for part of the 291 newly identified hot subdwarf stars in this study\footnote{This is only a subset of the table, the whole data will available online at the CDS ({\color{blue} https://cds.unistra.fr}) after publication.}.}
\tabletypesize{\normalem}
\tablehead{
\colhead{obs\_id} &  
\colhead{SNRU} &\colhead{source\_id} & 
\colhead{$T_\mathrm{eff}$} & \colhead{ $\mathrm{log}\ g$} & 
\colhead{ $\log(n{\rm He}/n{\rm H})$} & \colhead{RV} &\colhead{spclass}\\ 
 \colhead{LAMOST} &  \colhead{LAMOST} & \colhead{\textit{Gaia}} & \colhead{(\textit{K})} & \colhead{(cm\ S$^{-2}$)} & \colhead{} & \colhead{(km\ S$^{-1}$)} &  \colhead{}}
\colnumbers
\startdata
14510055 & 11.0 & 3446390992217234304 & 22513$\pm$3219 & 5.38$\pm$0.32 & -0.71$\pm$0.42 & 30.0$^{+43.1}_{-23.0}$ & sdOB\\
56701174 & 17.0 & 289425824464565632 & 26363$\pm$1095 & 5.1$\pm$0.06 & -0.87$\pm$0.16 & 30.0$^{+12.2}_{-11.4}$ & sdOB\\
74415176 & 15.0 & 2679569936368095488 & 62690$\pm$1304 & 6.28$\pm$0.02 & -1.63$\pm$0.3 & 470.0$^{+27.6}_{-1.2}$ & sdO\\
81504147 & 15.0 & 3381709575012174720 & 58457$\pm$1354 & 6.23$\pm$0.05 & -3.36$\pm$0.32 & 650.0$^{+18.3}_{-25.9}$ & sdO\\
115709085 & 24.0 & 3642680992030225536 & 35821$\pm$618 & 5.86$\pm$0.08 & -1.66$\pm$0.09 & -80.0$^{+9.9}_{-4.6}$ & sdOB\\
141110222 & 9.0 & 3833693825659533056 & 64607$\pm$4213 & 6.02$\pm$0.21 & -0.48$\pm$0.73 & 47.8$^{+66.3}_{-69.2}$ & sdO\\
164306246 & 73.0 & 1914301803258669440 & 27620$\pm$253 & 5.46$\pm$0.03 & -2.37$\pm$0.05 & 0.0$^{+2.8}_{-0.0}$ & sdB\\
184210031 & 15.0 & 364729314267240192 & 74644$\pm$3319 & 6.39$\pm$0.1 & -0.85$\pm$0.62 & -334.2$^{+20.7}_{-3.7}$ & sdO\\
188009108 & 13.0 & 217699973701423872 & 27385$\pm$976 & 5.26$\pm$0.11 & -2.43$\pm$0.14 & 80.0$^{+7.8}_{-1.5}$ & sdB\\
188205168 & 9.0 & 793520462344141056 & 63706$\pm$3485 & 6.43$\pm$0.1 & -1.94$\pm$0.42 & -60.0$^{+57.2}_{-47.9}$ & sdO\\
255801175 & 9.0 & 2863852345791113088 & 30995$\pm$1275 & 5.66$\pm$0.2 & -2.51$\pm$0.21 & 30.0$^{+9.2}_{-14.1}$ & sdB\\
256711147 & 22.0 & 389835459698932096 & 29251$\pm$577 & 5.33$\pm$0.08 & -3.39$\pm$0.33 & -47.7$^{+6.9}_{-2.3}$ & sdB\\
271004221 & 85.0 & 384468910944036992 & 28829$\pm$217 & 5.44$\pm$0.02 & -2.53$\pm$0.03 & -100.0$^{+0.0}_{-0.0}$ & sdB\\
301309205 & 24.0 & 218951870771462144 & 78887$\pm$2384 & 6.94$\pm$0.08 & -1.55$\pm$0.18 & 90.0$^{+19.5}_{-16.3}$ & sdO\\
302705206 & 10.0 & 198847055842102144 & 29225$\pm$1178 & 5.48$\pm$0.13 & -1.77$\pm$0.09 & 56.1$^{+5.2}_{-3.4}$ & sdOB\\
335916186 & 18.0 & 1185738013981539840 & 55507$\pm$1784 & 6.48$\pm$0.05 & -3.18$\pm$0.21 & -30.0$^{+7.6}_{-24.9}$ & sdO\\
343902066 & 43.0 & 1519860699806445184 & 66932$\pm$1566 & 6.88$\pm$0.05 & 0.55$\pm$0.22 & -510.9$^{+12.2}_{-14.6}$ & sdO\\
347405022 & 13.0 & 4581440480773963264 & 64997$\pm$4423 & 5.88$\pm$0.07 & -1.98$\pm$0.24 & -160.0$^{+22.7}_{-5.0}$ & sdO\\
401216228 & 26.0 & 3817717887347994112 & 31484$\pm$380 & 5.9$\pm$0.06 & -2.51$\pm$0.08 & 0.0$^{+1.2}_{-1.2}$ & sdB\\
405902229 & 47.0 & 300394067131824768 & 77818$\pm$2912 & 6.98$\pm$0.15 & -1.72$\pm$0.25 & 50.0$^{+17.2}_{-7.9}$ & sdO\\
422904134 & 8.0 & 957591309724015872 & 62848$\pm$5166 & 5.81$\pm$0.25 & 0.12$\pm$0.8 & -147.8$^{+148.3}_{-58.2}$ & He-sdO\\
451310021 & 10.0 & 3725037883284023680 & 64511$\pm$4074 & 6.25$\pm$0.18 & -0.81$\pm$0.77 & -50.0$^{+52.4}_{-21.3}$ & sdO\\
474506103 & 15.0 & 3149568478152254208 & 53418$\pm$3829 & 5.61$\pm$0.11 & -0.94$\pm$0.42 & -230.0$^{+1.1}_{-23.9}$ & He-sdO\\
491112208 & 10.0 & 397750706468048896 & 26706$\pm$2004 & 5.31$\pm$0.19 & -2.05$\pm$0.27 & 30.9$^{+26.3}_{-10.7}$ & He-sdB\\
498302133 & 17.0 & 974594466774413440 & 32435$\pm$1131 & 5.21$\pm$0.1 & -1.77$\pm$0.21 & 24.3$^{+4.8}_{-4.3}$ & sdOB\\
499501249 & 9.0 & 194449627807985920 & 25316$\pm$1268 & 5.45$\pm$0.15 & -1.75$\pm$0.22 & -70.0$^{+6.8}_{-5.4}$ & sdB\\
532409153 & 9.0 & 1033476235018600320 & 30107$\pm$2562 & 5.5$\pm$0.24 & -1.87$\pm$0.31 & 0.0$^{+27.4}_{-27.9}$ & sdB\\
537915156 & 9.0 & 3328166141880416512 & 35870$\pm$985 & 5.62$\pm$0.1 & -2.88$\pm$0.33 & 70.0$^{+7.3}_{-9.9}$ & sdB\\
547010211 & 10.0 & 815981251516846080 & 37282$\pm$3425 & 5.71$\pm$0.18 & -1.1$\pm$0.35 & 255.2$^{+4.8}_{-2.0}$ & sdOB\\
555614008 & 12.0 & 3959631234670040704 & 34537$\pm$1052 & 6.13$\pm$0.08 & -1.68$\pm$0.18 & -11.4$^{+12.7}_{-12.0}$ & sdOB\\
566110031 & 46.0 & 1234828283288291840 & 36435$\pm$478 & 6.37$\pm$0.04 & -1.73$\pm$0.1 & -30.0$^{+0.0}_{-0.0}$ & sdO\\
587411165 & 9.0 & 1938255660503776000 & 62926$\pm$4118 & 6.02$\pm$0.21 & -0.27$\pm$0.79 & -143.9$^{+98.2}_{-78.9}$ & sdO\\
592501071 & 31.0 & 1746866317154525696 & 27751$\pm$390 & 5.47$\pm$0.04 & -3.03$\pm$0.12 & -20.0$^{+1.0}_{-1.1}$ & sdB\\
593805115 & 20.0 & 1964027697669073408 & 49239$\pm$2221 & 5.64$\pm$0.1 & -2.57$\pm$0.25 & -110.9$^{+17.1}_{-15.6}$ & sdO\\
593816107 & 9.0 & 1965596391525733120 & 28970$\pm$1474 & 5.51$\pm$0.16 & -2.26$\pm$0.17 & -70.0$^{+11.2}_{-6.5}$ & sdB\\
593816248 & 8.0 & 1968673301791825920 & 28203$\pm$1381 & 5.48$\pm$0.14 & -2.67$\pm$0.17 & -62.6$^{+6.2}_{-14.0}$ & sdB\\
594011103 & 30.0 & 391755859836391680 & 28403$\pm$353 & 5.47$\pm$0.05 & -3.11$\pm$0.18 & -80.0$^{+3.8}_{-1.2}$ & sdB\\
601906006 & 15.0 & 268227442145432960 & 38156$\pm$1164 & 5.88$\pm$0.08 & -1.78$\pm$0.08 & -30.9$^{+0.9}_{-5.4}$ & sdOB\\
604814236 & 8.0 & 3225258996047056768 & 27988$\pm$2438 & 5.55$\pm$0.22 & -1.8$\pm$0.33 & 117.3$^{+2.7}_{-2.2}$ & sdB\\
605216183 & 15.0 & 1880916403991110144 & 21726$\pm$515 & 5.13$\pm$0.05 & -2.11$\pm$0.12 & -130.0$^{+4.0}_{-0.0}$ & sdB\\
632802159 & 8.0 & 671999108644503040 & 21754$\pm$901 & 5.25$\pm$0.12 & -2.06$\pm$0.19 & 10.0$^{+8.0}_{-4.5}$ & sdB\\
645909165 & 23.0 & 3425246382885368576 & 26690$\pm$676 & 5.26$\pm$0.08 & -2.99$\pm$0.15 & 33.6$^{+3.6}_{-3.4}$ & sdB\\
663709117 & 16.0 & 4459185168703405568 & 51291$\pm$1345 & 6.5$\pm$0.03 & -2.81$\pm$0.17 & 80.0$^{+18.4}_{-4.5}$ & sdO\\
664309155 & 8.0 & 4540317135283890048 & 21223$\pm$1708 & 5.15$\pm$0.13 & -0.81$\pm$0.32 & -391.1$^{+13.3}_{-24.8}$ & sdB\\
\enddata
\end{deluxetable}

During data processing, we employed the Monte Carlo method to estimate uncertainties in atmospheric parameters. Specifically, for each observed spectrum, we generated 100 synthetic spectra by drawing flux values at each wavelength sampling point from a Gaussian distribution. This distribution used the observed flux value as its mean and the corresponding flux uncertainty as its standard deviation. These 100 synthetic spectra were then individually processed through our trained deep learning model to predict their atmospheric parameters. Subsequently, we fitted Gaussian functions to the resulting distributions of the three atmospheric parameters (i.e., $T_{\rm eff}$, ${\rm log}\ g $ and $\log(n{\rm He}/n{\rm H})$), respectively. The mean of each Gaussian fit represents the final atmospheric parameter value for the original observed spectrum, while the standard deviation of the fit quantifies its uncertainty. 

\begin{figure}
    \centering    
    
    \includegraphics[width=85mm]{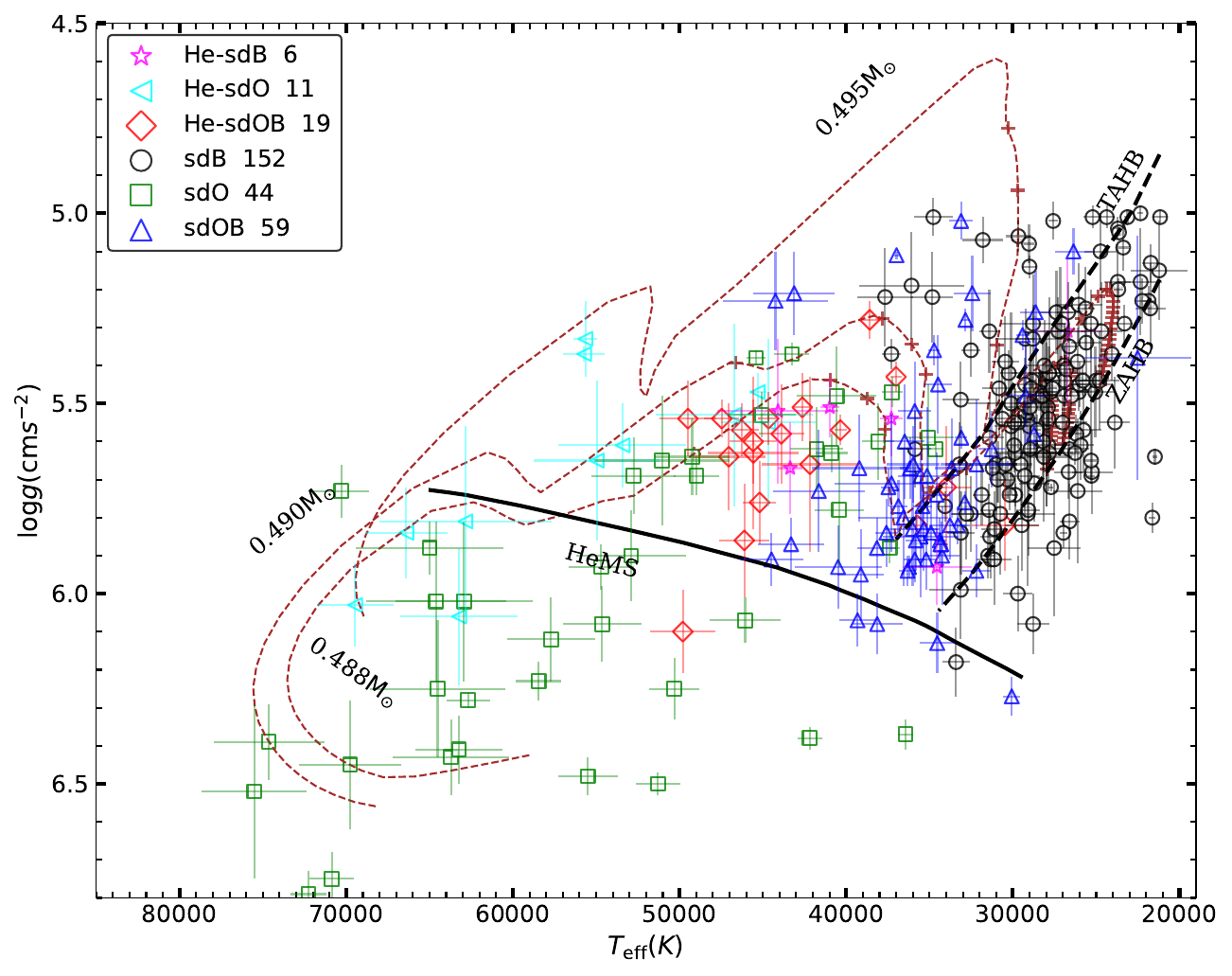}
    \includegraphics[width=85mm]{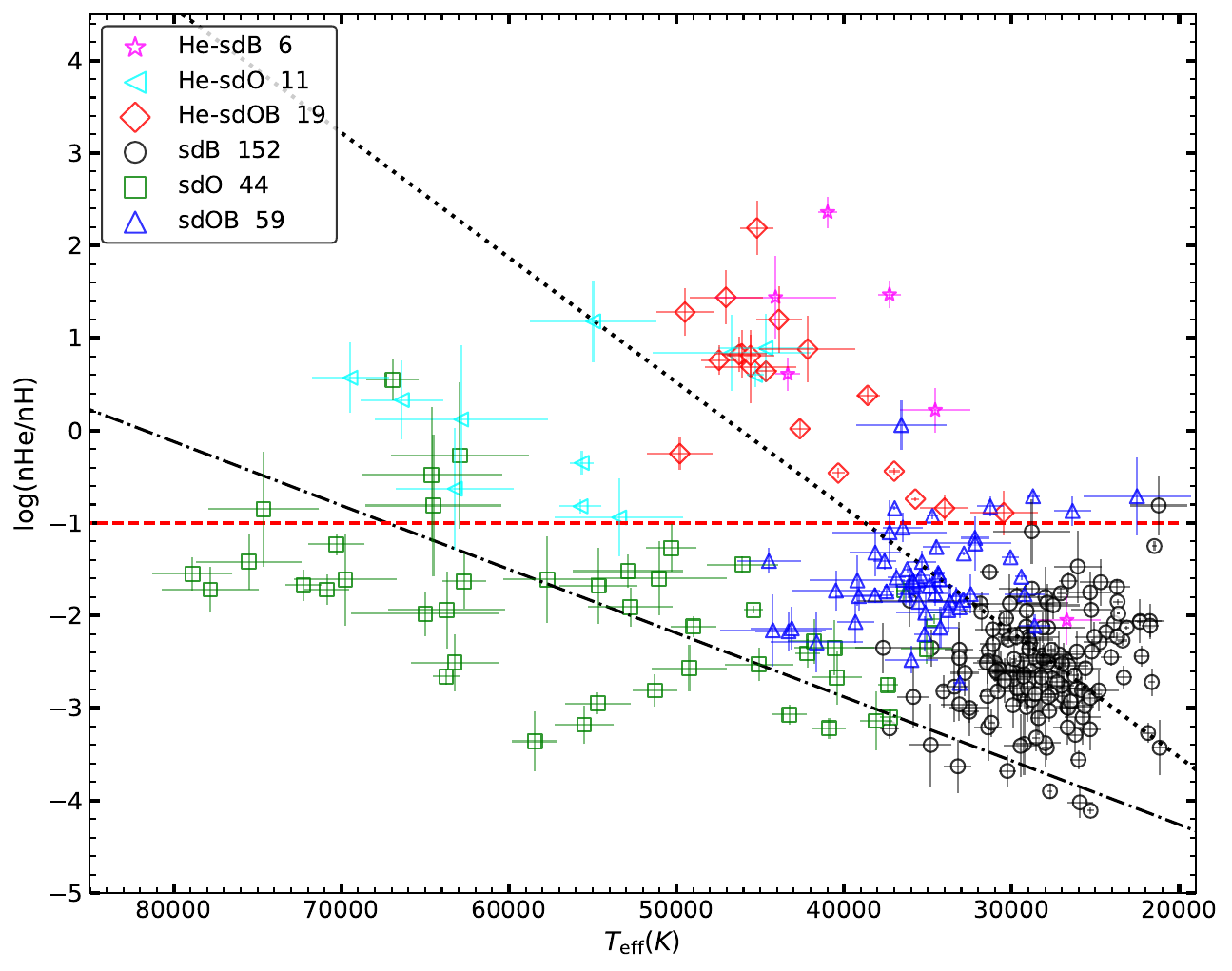}
    \includegraphics[width=85mm]{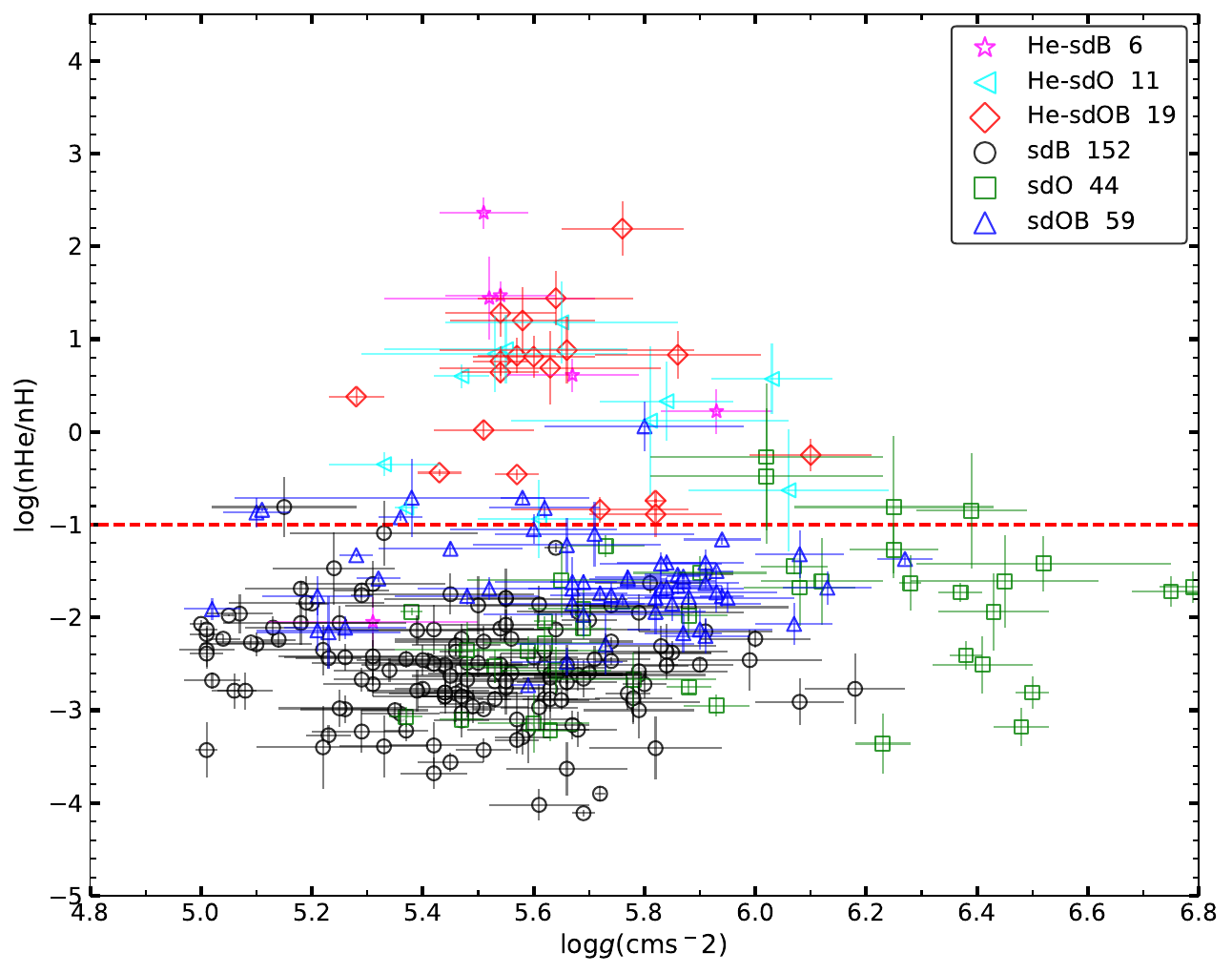}
    \caption{The relationships among atmospheric parameters for the 291 newly identified hot subdwarf stars. The markers and number counts for different types of hot subdwarfs are shown in each panel. Top left panel: $T_{\rm eff}$ vs $\log{g}$ plane.
    The ZAHB and TAHB sequences with [Fe/H]= -1.48 from \citet{1993ApJ...419..596D} are denoted by dashed lines. The He main-sequence from \citet{1971AcA....21....1P} is marked by black solid line. Three evolution tracks for hot HB stars from \citet{1993ApJ...419..596D} are presented by brown dotted curves, for which the masses from top to bottom are 0.495, 0.490, and 0.488 $M_{\odot}$, respectively. Top right panel: $T_{\rm eff}$-$\log(n{\rm He}/n{\rm H})$ plane. The black dotted line and dot-dashed line are the linear regression lines fitted for the two He sequences by \citet{2003A&A...400..939E} and \citet{2012MNRAS.427.2180N}, respectively. Bottom middle panel: $\log{g}$-$\log(n{\rm He}/n{\rm H})$. The red horizontal dashed line denotes the solar value of He abundance (e.g., $\log(n{\rm He}/n{\rm H})$ = -1)  }. 
    \label{fig new hot subdwarfs}
\end{figure}

Ultimately, 1512 sources were confirmed as hot subdwarfs, which have $T_{\rm eff} \geq$ 20\ 000 K and ${\rm log}\ g \geq$ 5.0.‌ However, among these 1512 hot subdwarfs, 1221 were previously identified and listed in the catalog of \citet{2022A&A...662A..40C}. It means we have newly confirmed 291 hot subdwarfs from Culpan's candidate catalog.‌ We also applied the spectral classification method from \citet{2018ApJ...868...70L} to the 291 newly confirmed hot subdwarfs (also see \citealt{1990A&AS...86...53M, 2017A&A...600A..50G}) and obtained their spectral classifications. This sample contains 152 sdB stars, 59 sdOB stars, 44 sdO stars, 19 He-sdOB stars, 11 He-sdO stars, and 6 He-sdB stars. All newly confirmed hot subdwarfs are cataloged in Table 2.‌ 

Table 2 lists the main parameters of 291 newly confirmed hot subdwarfs identified via deep learning models. From left to right, columns 1-2 provide the LAMOST obsid and SNRu. Columns 3 give the Gaia EDR3 source\_id. Columns 4-7 present the atmospheric parameters and their uncertainties, along with radial velocity and its uncertainty, which are derived in this study. The last column indicates the spectral classification of the hot subdwarfs. 

Figure \ref{fig new hot subdwarfs} presents correlations among atmospheric parameters for the 291 newly identified hot subdwarfs across three panels: $T_{\rm eff}$ vs   ${\rm log}\ g$ (top left panel), $T_{\rm eff}$ vs $\rm log(n{\rm He}/n{\rm H})$ (top right panel) and ${\rm log}\ g$ vs $\rm log(n{\rm He}/n{\rm H})$ (bottom middle panel). The distributions confirm that the new sample exhibits consistent characteristics with previously known hot subdwarfs \citep{2019ApJ...881..135L}. Most of sdB and sdOB stars reside between the zero-age horizontal branch (ZAHB) and terminal-age horizontal branch (TAHB), indicating ongoing ‌core helium burning. Higher temperature sdOs appear to have completed core helium burning and are likely ‌evolving toward the WD cooling curve. Distinct ‌He-poor‌ and ‌He-rich‌ sequence are clearly manifested in the $T_{\rm eff}$ vs $\log(n{\rm He}/n{\rm H})$ plane, which is matching prior findings. 

\section{Discussion and summary} 
In this study, we trained deep learning models on tens of thousands of synthetic and observed spectra of hot subdwarfs, resulting in nine distinct models tailored to LAMOST spectra with different SNRs. These models could be respectively applicable to LAMOST spectra across varying SNR ranges to predict atmospheric parameters for hot subdwarf stars. The models demonstrated outstanding performance on the test set, with coefficients of determination ($R^{2}$) all exceeding 0.96. When applied to confirmed hot subdwarfs, the models also achieved highly accurate atmospheric parameters (see Fig \ref{fig comparison with culpan}), with MAE of 730 K for $T_{\rm eff}$, 0.09 dex for ${\rm log}\ g$, and 0.03 dex for $\log(n{\rm He}/n{\rm H})$, respectively.  For comparison, we cross-matched confirmed hot subdwarfs from \citet{2018ApJ...868...70L, 2019ApJ...881..135L,2020ApJ...889..117L} with those in \citet{2021ApJS..256...28L}, calculating the MAE in atmospheric parameters for the 733  common objects, and obtained value of 1070 K for $T_{\rm eff}$, 0.07  dex for ${\rm log}\ g$, and 0.07 dex for $\log(n{\rm He}/n{\rm H})$, respectively.‌  These results demonstrate that our trained deep learning models exhibit excellent generalization capabilities. Their accuracy in predicting atmospheric parameters of hot subdwarfs is  comparable to traditional methods, or even better (e.g., for $T_{\rm eff}$ and $\log(n{\rm He}/n{\rm H})$). 

While artificial intelligence methods have made notable progress in searching for hot subdwarfs, research employing such approaches to predict their atmospheric parameters remains scarce. \citet{2024ApJS..274....2C} utilized a Se-ResNet + SVM model to select 3086 hot subdwarf candidates within LAMOST DR8, subsequently confirming 168 newly discovered hot subdwarfs. Additionally, they employed the Se-ResNet algorithm to predict atmospheric parameters of hot subdwarfs. Their training sample consisted of only 1858 LAMOST spectra with known atmospheric parameters. Due to this  limited training sample size, their model exhibited large mean absolute errors (MAEs) in the test set: 3009 K for $T_{\rm eff}$, 0.2 dex for ${\rm log}\ g$, and 0.42 dex for $\log(n{\rm He}/n{\rm H})$ (see Section 4.3 and Fig 7 in their study). These errors substantially exceed those achieved by this study, and are also considerably larger than errors obtained via traditional spectral fitting methods. 

Although our deep learning model has achieved excellent overall performance in predicting the atmospheric parameters of hot subdwarfs, some minor issues remain. For instance, predictions for extremely He-poor and extremely He-rich stars still pose challenges. Predicting He abundance for the test set spectra with SNRu = 25 (see top right panel of Fig \ref{fig comparsion for test data}) reveals that our predictions are a little higher for abundances below -4.0 and lower for abundances above 1.5. Although this phenomenon is less pronounced in the case of SNRu=80 (bottom right panel of Fig \ref{fig comparsion for test data}), it still persists.
This difficulty arises because the spectral features for hot subdwarfs with extreme He deficiency or richness are very similar. For example, spectra with $\log(n{\rm He}/n{\rm H})$ $\leq$ -4.0 exhibit virtually no He lines, being dominated almost entirely by H Balmer lines. This makes it exceptionally difficult to distinguish between them in the spectra, hindering ability of the model to capture the subtle spectral variations. It should be noted that traditional spectral fitting methods also face similar difficulties when processing low-resolution spectra of stars with extremely He-poor or He-rich spectra. 

In each group of experimental data, there are 11 396 synthetic spectra at a specific SNR and 945 observed spectra at varying SNRs. A potentially better approach might be to classify the 945 observed spectra according to their SNR and then assign observed spectra to the group of synthetic spectra with same SNR as the observed spectra. However, due to the limited number of observed spectra, this would cause a problem that certain atmospheric parameter values in the training samples would lack corresponding observed spectra, having only synthetic spectra. This absence would negatively impact the final results. Conversely, we added all 945 observed spectra into each specific-SNR synthetic spectra set, because there are only a few observed spectra with the same or similar SNR for certain atmospheric parameters, their impacts on the final results are not decisive.
Furthermore, as discussed in Section 2.3, the atmospheric parameters of some hot subdwarfs exceed the coverage of our synthetic spectra library (e.g., $T_{\rm eff} > 56\ 000$K, ${\rm log}\ g> 6.3$). The inclusion of observed  spectra effectively compensates for this gap by providing training samples with higher temperatures and surface gravities. 
In this case, it's primary role is to act as a calibration or adjustment to the predictions derived from the synthetic spectra. Actually, we also have  experimented with training using only the 11 396 synthetic spectra, but the final results proved less effective than the current approach that incorporates the observed spectra.

\begin{figure}
    \centering    
    
    \includegraphics[width=85mm]{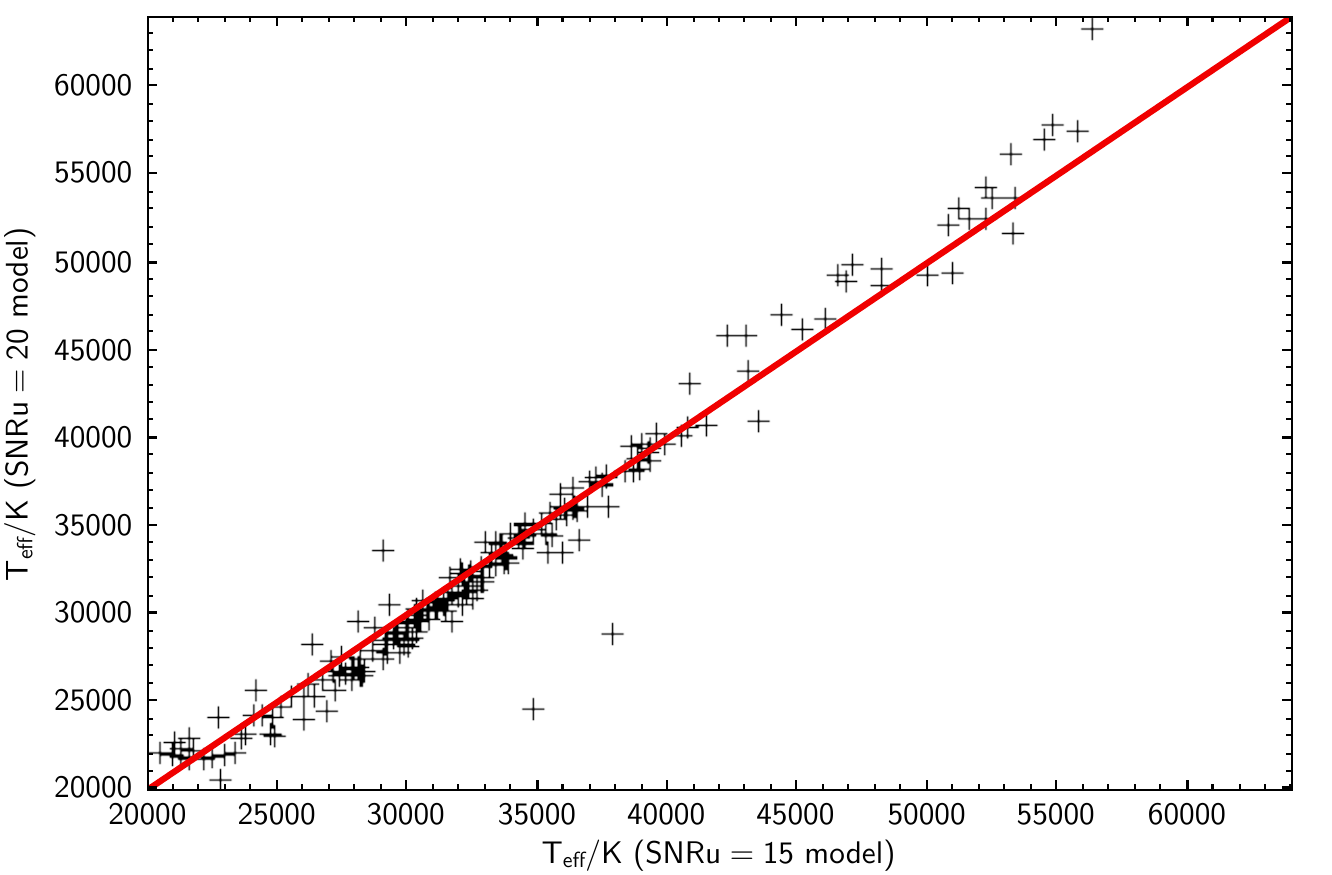}
    \includegraphics[width=85mm]{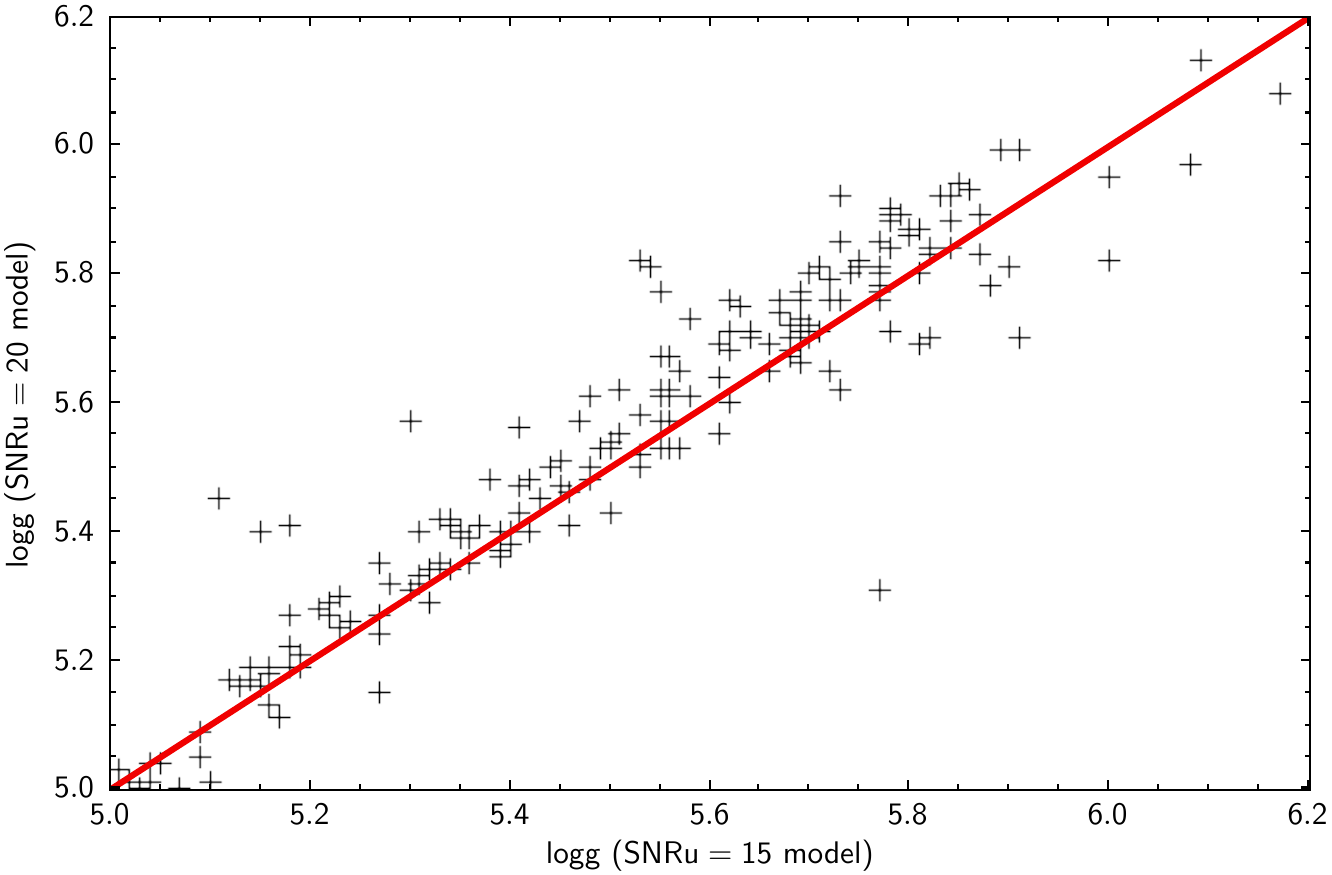}
    \includegraphics[width=85mm]{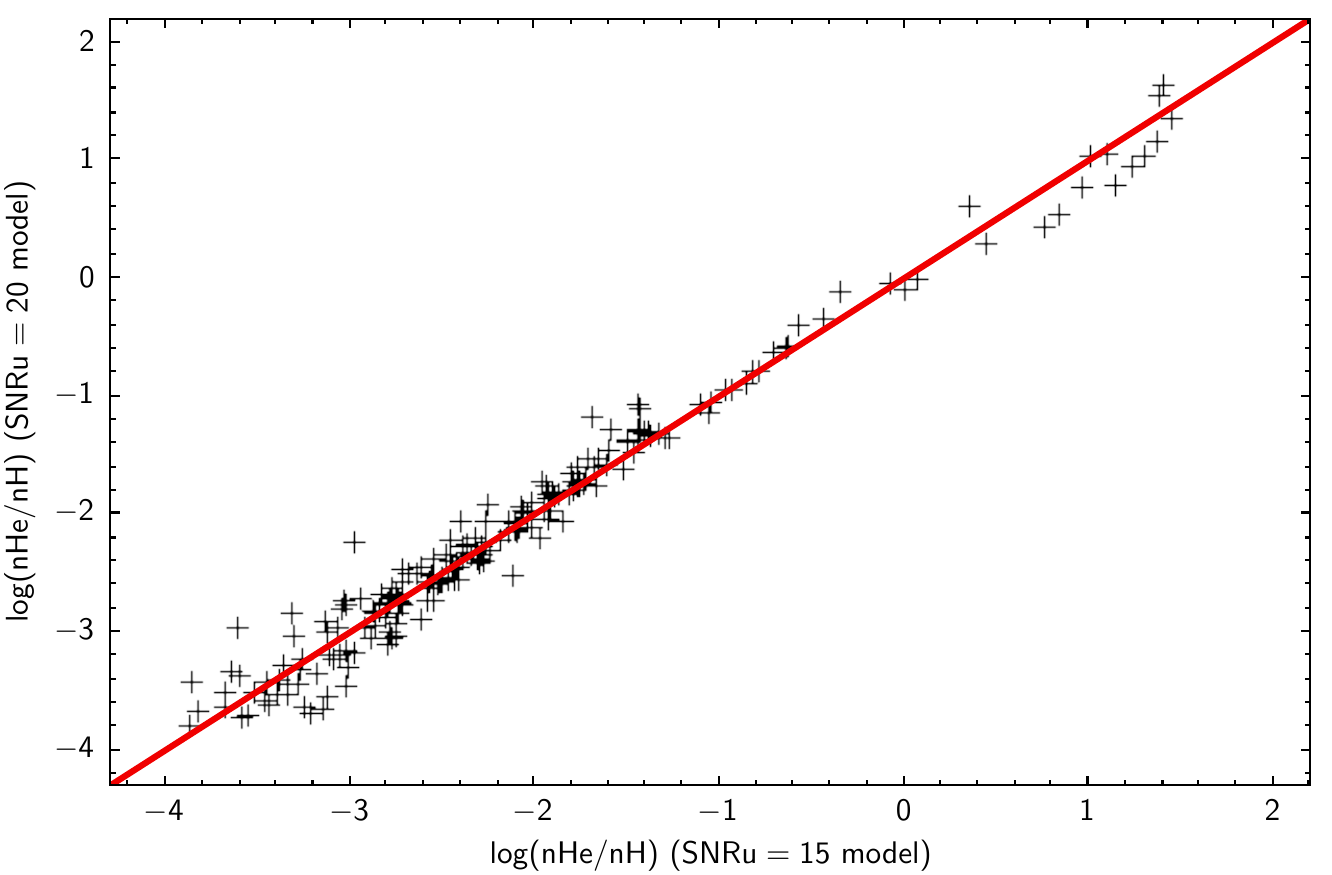}
    \caption{ Comparison of atmospheric parameters predicted by SNRu=15 and SNRu=20 models for observed  spectra with SNRu between 15 and 20.
    ‌Left top panel:‌ $T_{\rm eff}$ comparison.
    ‌Right top panel:‌ ${\rm log}\ g$ comparison.
    ‌Bottom middle panel:‌ $\log(n{\rm He}/n{\rm H})$  comparison. In all panels, the horizontal axis represents the predicted values from the ‌SNRu = 15 model‌. While the vertical axis represents the predicted values from the ‌SNRu = 20 model‌. The MAE are  $\Delta T_{\rm eff}=$ 241 K, $\Delta {\rm log}\ g=$ 0.04 dex, and $\Delta \log(n{\rm He}/n{\rm H})=$ 0.02 dex, respectively }. 
    \label{fig snru15_vs_snru20} 
\end{figure}

To test the dependence of the training models on  SNR, we compared the prediction results for observed spectra with SNRu between 15 and 20, using models at SNRu = 15 and 20, respectively. Figure \ref{fig snru15_vs_snru20} shows diagonal plots comparing the three atmospheric parameters ($T_{\rm eff}$, ${\rm log}\ g$, and $\log(n{\rm He}/n{\rm H})$) predicted by the SNRu = 15 model (horizontal axis) and SNRu = 20 model (vertical axis). Despite the SNRu value  for the observed spectra span of 5, the three parameters show excellent agreement. The mean absolute errors (MAE)  are  $\Delta T_{\rm eff}=$ 241 K, $\Delta {\rm log}\ g=$ 0.04 dex, and $\Delta \log(n{\rm He}/n{\rm H})=$ 0.02 dex, respectively. A similar comparison for observed spectra with 45 $\le$\ SNRu $\le $ 60  using models at SNRu=45 and 60 also yielded consistent results, even with the larger SNRu span of 15. The MAEs for this set are $\Delta T_{\rm eff}=$ 236 K, $\Delta {\rm log}\ g=$ 0.02 dex, and $\Delta \log(n{\rm He}/n{\rm H})=$ 0.04 dex, respectively. Therefore, small changes in spectral SNRu have a very limited influence on the model's predictions, with an even smaller effect for higher SNRu spectra.

In this paper, the model we trained is for LAMOST spectra, because the resolution of synthetic spectra is reduced by convolution  to match the low - resolution spectra of LAMOST. If it is to be applied to other large - scale survey spectra, synthetic spectra with corresponding resolutions need to be regenerated according to their resolutions for training. In fact, we have used a similar method to apply the trained model to DESI DR1 spectra to identify hot subdwarfs (Dong et al. in preparation). In the future, we plan to organize these models and release them on GitHub (\textcolor{blue}{www.github.com}) so that other scholars can also use the models to obtain parameters of hot subdwarfs.

Overall, after incorporating both synthetic and observational spectra into the training dataset, our deep learning models have not only achieved accuracy comparable to traditional methods in predicting atmospheric parameters for hot subdwarfs,  but also far surpassed them in efficiency and speed. For instance, traditional spectral fitting methods require 5-30 minutes to converge and obtain the atmospheric parameters for a single spectrum. In contrast, our trained deep learning models can predict the atmospheric parameters for thousands of spectra within tens of seconds to several minutes. 
Furthermore, traditional spectral fitting methods struggle significantly with analyzing extremely low-resolution spectra (e.g., $\frac{\lambda}{\Delta \lambda}$ $\le$ 200). The deep learning model, however, is expected to be capable of making predictions for spectra with even lower resolutions, such as those from Gaia XP spectra \citep{2023A&A...674A...1G} and the upcoming China Space Station Telescope (CSST, \citealt{2011SSPMA..41.1441Z}) mission. We will progressively complete and advance these developments in the near future. This will increase the number of confirmed hot subdwarfs with accurately determined atmospheric parameters by an order of magnitude. 

\begin{acknowledgments}
We thank the anonymous referee for the valuable suggestions and comments that helped improve the manuscript greatly. This work acknowledges support from the National Natural Science Foundation of China (Nos. 12073020, 12588202 and 12273055), Scientific Research Fund of Hunan Provincial Education Department grant No. 20K124, the National Key R\&D Program of China, grant no. 2023YFE0107800 and 2024YFA1611902,
the International Partnership Program of Chinese Academy of Sciences under grant No. 178GJHZ2022040GC. JZ acknowledge the surport by the National Astronomical Observatories , CAS grant no. E4ZB0301 and the support from the Strategic Priority Research Program of Chinese Academy of Sciences, grant No. XDB1160301. K.H. acknowledges support from the Scientific Research Funds of Hunan Provincial Education Department (Nos. 22A0099 and 24A0101). Guoshoujing Telescope (the Large Sky Area Multi-Object Fiber Spectroscopic Telescope LAMOST) is a National Major Scientific Project built by the Chinese Academy of Sciences. Funding for the project has been provided by the National Development and Reform Commission. LAMOST is operated and managed by the National Astronomical Observatories, Chinese Academy of Sciences.
\end{acknowledgments}

\software{astropy \citep{2013A&A...558A..33A,2018AJ....156..123A}, TOPCAT \citep{2005ASPC..347...29T}, Tensorflow \citep{abadi2016tensorflowlargescalemachinelearning}, LASPEC \citep{2020ApJS..246....9Z,2021ApJS..256...14Z}, Python \ ({\textcolor{blue}{www.python.org}})}


\bibliography{sd_parameter_2025}{}
\bibliographystyle{aasjournalv7}



\end{document}